\documentclass{article} 
\usepackage{iclr2023_conference,times}

\iclrfinalcopy

\usepackage{amssymb,amsmath,enumerate,threeparttable,bm,hyperref}
\usepackage{caption}
\usepackage{subfigure}
\usepackage{algorithm}
\usepackage{algorithmic}
\usepackage{booktabs}
\usepackage{multirow}
\usepackage{colortbl}
\usepackage{threeparttable}
\usepackage{color}

\usepackage{lipsum}
\usepackage{enumitem}

\usepackage{setspace}
\usepackage{amsmath}
\usepackage{wrapfig}
\usepackage[export]{adjustbox}
\usepackage[switch]{lineno}

\usepackage{array}
\newcommand{\tabincell}[2]{\begin{tabular}{@{}l#1@{}}#2\end{tabular}}

\usepackage{listings}
\usepackage{xcolor}
\definecolor{dkgreen}{rgb}{0,0.6,0}
\definecolor{gray}{rgb}{0.5,0.5,0.5}
\definecolor{mauve}{rgb}{0.58,0,0.82}

\title{BackdoorBox: A Python Toolbox for\\Backdoor Learning}

\author{Yiming Li\thanks{The first two authors contributed equally to this toolbox.} , Mengxi Ya$^{\ast}$, Yang Bai, Yong Jiang, Shu-Tao Xia\\
Tsinghua Shenzhen International Graduate School, Tsinghua University, China\\
\{li-ym18, yamx21, y-bai17\}@mails.tsinghua.edu.cn; \{jiangy, xiast\}@sz.tsinghua.edu.cn
}

\begin{document}

\maketitle

\begin{abstract}
Third-party resources ($e.g.$, samples, backbones, and pre-trained models) are usually involved in the training of deep neural networks (DNNs), which brings backdoor attacks as a new training-phase threat. In general, backdoor attackers intend to implant hidden backdoor in DNNs, so that the attacked DNNs behave normally on benign samples whereas their predictions will be maliciously changed to a pre-defined target label if hidden backdoors are activated by attacker-specified trigger patterns. To facilitate the research and development of more secure training schemes and defenses, we design an open-sourced Python toolbox that implements representative and advanced backdoor attacks and defenses under a unified and flexible framework. Our toolbox has four important and promising characteristics, including consistency, simplicity, flexibility, and co-development. It allows researchers and developers to easily implement and compare different methods on benchmark or their local datasets. This Python toolbox, namely \texttt{BackdoorBox}, is available at \url{https://github.com/THUYimingLi/BackdoorBox}.
\end{abstract}

\section{Introduction}
Deep neural networks (DNNs) have demonstrated their superiority in computer vision \citep{lecun2015deep,li2015learning,qiu2021end2end}. In general, training a well-performed model requires a large number of samples and computational resources. Accordingly, researchers and developers often use third-party resources ($e.g.$, samples and backbones) during the training process or even directly adopt third-party pre-trained models for convenience. However, the training opacity also introduces the backdoor threat \citep{gao2020backdoor,goldblum2022dataset,li2022backdoor}. Specifically, the backdoor adversaries can implants hidden backdoor into DNNs by maliciously manipulating the training process ($e.g.$, samples or loss). The attacked models behave normally in predicting benign samples while having abnormal behaviors when the backdoor is activated by poisoned samples.

Currently, there were many existing backdoor attacks and defenses \citep{lin2020composite,xiang2021backdoor,zhai2021backdoor,shen2021backdoor,tao2022better,xiang2022post}. Although most of them were open-sourced, there is still no toolbox that can easily and flexibly implement and compare them simultaneously. To fill this gap, we design and develop \texttt{BackdoorBox} — a comprehensive open-sourced Python toolbox that implements representative and advanced backdoor attacks and defenses under a \emph{unified} framework that can be used in a \emph{flexible} manner.

\begin{table}[ht]
\centering
\caption{Implemented backdoor attacks in \texttt{BackdoorBox}.}
\vspace{-1em}
\scalebox{1}{
\begin{tabular}{lll}
\toprule
\textbf{Adversary's Capacity}                 & \textbf{Attack Method}                  & \textbf{Key Properties}                          \\ \midrule
\multirow{9}{*}{Poison-only}         & BadNets \citep{gu2019badnets}               & Visible                                 \\ \cline{2-3}
                                     & Blended \citep{chen2017targeted}         & Invisible                               \\ \cline{2-3}
                                     & WaNet \citep{nguyen2021wanet}                   & Invisible              \\ \cline{2-3}
                                     & ISSBA \citep{li2021invisible}                  & \tabincell{l}{Invisible\\ Sample-specific\\ Physical}    \\ \cline{2-3}
                                     & Label-consistent \citep{turner2019label} & \tabincell{l}{Invisible\\ Clean-label}                  \\ \cline{2-3}
                                     & TUAP \citep{zhao2020clean}                   & \tabincell{l}{Invisible\\ Clean-label}                  \\ \cline{2-3}
                                     & Refool \citep{liu2020reflection}                 & \tabincell{l}{Sample-specific\\ Clean-label}            \\ \cline{2-3}
                                     & Sleeper Agent \citep{souri2022sleeper}          & \tabincell{l}{Invisible\\ Sample-specific\\ Clean-label} \\ \cline{2-3} 
                                     & UBW-P \citep{li2022untargeted}          & \tabincell{l}{Untargeted\\Dispersable} \\ \hline
\multirow{4}{*}{Training-controlled} & Blind \citep{bagdasaryan2020blind}        & Non-optimized                           \\ \cline{2-3}
                                     & IAD \citep{nguyen2020input}                    & \tabincell{l}{Optimized\\ Sample-specific}              \\ \cline{2-3}
                                     & PhysicalBA \citep{li2021backdoor}         & Physical                                \\ \cline{2-3}
                                     & LIRA \citep{doan2021lira}                   & \tabincell{l}{Invisible\\ Optimized\\ Sample-specific} \\ \bottomrule
\end{tabular}
}
\label{tab:attack}
\end{table}

Compared to existing backdoor-related libraries ($e.g.$, TrojanZoo \citep{pang2022trojanzoo} and BackdoorBench \citep{wubackdoorbench2022}), \texttt{BackdoorBox} has four distinct advantages: \textbf{(1)} It contains more than 20 representative backdoor attacks and defenses covering both classical and advanced methods. \textbf{(2)} It develops all methods under a unified framework with high consistency. \textbf{(3)} \texttt{BackdoorBox} allows using user-specified samples and models for attacks and defenses. \textbf{(4)} It allows using attack and defense modules jointly or separately. We hope that our toolbox can facilitate the research and development of more secure training schemes and defenses against backdoor threats.

\section{Toolbox Characteristics and Dependencies}
\subsection{Toolbox Characteristics}
\noindent \textbf{Consistency.} Instead of developing each method separately and organizing them simply, we re-implement all methods in a unified manner. Specifically, variables having the same function have a consistent name. Similar methods inherit the same `base class' for further development, have a unified workflow, and have the same core sub-functions ($e.g.$, \texttt{get\_poisoned\_dataset}).

\vspace{0.2em}
\noindent \textbf{Simplicity.} We provide code examples for each implemented backdoor attack and defense to explain how to use them, the definitions and default settings of all required attributes, and the necessary code comments. Users can easily implement and develop our toolbox.

\vspace{0.2em}
\noindent \textbf{Flexibility.} We allow users to easily obtain important intermediate outputs and components of each method ($e.g.$, poisoned dataset and attacked/repaired model), use their local samples and model structure for all implemented attacks and defenses, and interact with their local codes. The attack and defense modules can be used jointly or separately.

\vspace{0.2em}
\noindent \textbf{Co-development.} All codes and developments of \texttt{BackdoorBox} are hosted on \href{https://github.com/THUYimingLi/BackdoorBox}{GitHub} to facilitate collaboration. At the time of this writing, there are more than seven contributors have helped develop the code base and others have contributed to the code test. The co-development mode facilitates rapid and comprehensive developments and bug findings.

\subsection{Toolbox Dependencies} 
Currently, our \texttt{BackdoorBox} is only compatible with Python 3 using PyTorch. It also relies on \texttt{numpy}, \texttt{scipy}, \texttt{opencv}, \texttt{pillow}, \texttt{matplotlib}, \texttt{requests}, \texttt{termcolor}, \texttt{easydict}, \texttt{seaborn}, \texttt{imageio}, and \texttt{lpips}. The full dependent package list is included in the \href{https://github.com/THUYimingLi/BackdoorBox/blob/main/requirements.txt}{`requirements.txt'}. People can easily download all required packages by running:  

\framebox[\linewidth]{pip install -r requirements.txt}

\section{The Module of Backdoor Attacks}
\label{sec:attacks}
\subsection{General Information}
We categorize existing backdoor attacks into three main types, including \textbf{(1)} poison-only backdoor attacks, \textbf{(2)} training-controlled backdoor attacks, and \textbf{(3)} model-modified backdoor attacks, based on the capacities of backdoor adversaries. Specifically, under the poison-only setting, the adversaries can only poison training samples while having no information and cannot control the training schedule \citep{gu2019badnets, li2021invisible, qi2023revisiting}; The training-controlled backdoor attacks \citep{zeng2021rethinking,cheng2021deep,zhao2022defeat} allow adversaries to fully control the whole training process, such as training samples and the training schedule; Model-modified attacks \citep{tang2020embarrassingly,qi2022towards,bai2022hardly} enable adversaries to directly modify the model by inserting malicious sub-modules or flipping its critical bits (usually in the deployment stage).

Currently, our toolbox has implemented nine poison-only attacks and four training-controlled attacks, as shown in Table \ref{tab:attack}. We have not implemented any model-modified attack at this time because these methods are usually non-poisoning-based, having limited threat scenarios and well-developed approaches. We will keep updating this toolbox to include more representative attacks.

\subsection{Library Design and Implementation}

\noindent \textbf{Design.} 
In our toolbox, since they enjoy a similar or even the same pipeline, all implemented poison-only backdoor attacks inherit from a \href{https://github.com/THUYimingLi/BackdoorBox/blob/main/core/attacks/base.py}{base class} with the same interface. Firstly, it first \texttt{check} whether the provided datasets are supported by our toolbox. Currently, our toolbox supports official MINIST and CIFAR10 datasets as well as arbitrary local datasets loaded by \texttt{torchvision.datasets.DatasetFolder}. After that, it will generate poisoned training and testing datasets that can be obtained by \texttt{get\_poisoned\_dataset}, based on which to \texttt{train} and \texttt{test} with the given schedule. During the training and testing process, it will calculate evaluation metrics, print running statements, and save necessary materials after every given interval. Users can also get the attacked model by \texttt{get\_model} when the training is finished. In particular, we implement the trigger appending process in the form of \texttt{torchvision.transforms}. More importantly, our toolbox allows adding this process to any particular place of the transformation sequence (instead of only at the end of it, as done by most of existing methods) by assigning the \texttt{poisoned\_transform\_train\_index} and the \texttt{poisoned\_transform\_test\_index} attributes. Accordingly, our toolbox is flexible, allowing users to simulate real backdoor scenarios. This is one of the advantages of our toolbox, compared to existing backdoor-related code bases ($e.g.$, TrojanZoo \citep{pang2022trojanzoo} and BackdoorBench \citep{wubackdoorbench2022}). For training-controlled attacks, some of them did not inherit from the previous base class, since they have different manners. However, we still unify the name of important variables, attributes, and functions for the convenience of users. This is also another distinctive advantage of our toolbox. In addition, we also implement a bool-type attribute \texttt{deterministic} for reproducing the results when it set to True.

\vspace{0.3em}
\noindent \textbf{Implementation.} To execute an attack, users need to assign the hyper-parameters of the (benign) training and testing datasets, the attack ($e.g.$, poisoning rate and target label), the training schedule ($e.g.$, model structure and learning rate), and the implementation ($e.g.$, adopted GPU and saving directory). Specifically, users should \textbf{(1)} load necessary packages, \textbf{(2)} load benign training and testing datasets, \textbf{(3)} define attack parameters, \textbf{(4)} initialize the attack, \textbf{(5)} assign training schedule, and \textbf{(6)} train (and obtain) the attacked model. The example of using BadNets is demonstrated in Appendix \ref{appendix:badnets_demo}. For training-controlled attacks ($e.g.$, IAD and PhysicalBA), as shown in Appendix \ref{appendix:physicalba_demo}, users should adopt \texttt{get\_poisoned\_dataset} after the \texttt{Attack.train} (instead of before it), since poisoned samples may most probably be updated during the training process. In particular, we have provided the example testing file for each attack in the \href{https://github.com/THUYimingLi/BackdoorBox/blob/main/tests}{tests} directory for reference.

\section{The Module of Backdoor Defenses}
\label{sec:defenses}

\subsection{General Information}
We categorize existing backdoor defenses into six main types, including \textbf{(1)} pre-processing-based defenses, \textbf{(2)} model repairing, \textbf{(3)} poison suppression, \textbf{(4)} model diagnosis, \textbf{(5)} sample diagnosis, and \textbf{(6)} certified defenses, based on defense properties. Specifically, the first type of method alleviates backdoor threats by pre-processing test images before feeding them into the (deployed) model for prediction \citep{liu2017neural,li2021backdoor, qiu2021deepsweep}, motivated by the observations that backdoor attacks may lose effectiveness when the trigger used for attacking is different from the one used for poisoning. These defenses are usually efficient and require minor defender capacities; Model repairing \citep{zhao2020bridging,wu2021adversarial,zeng2022adversarial} aims to erase potential hidden backdoor contained in given suspicious models; Poison suppression \citep{du2020robust,huang2022backdoor,wang2022training} intends to depress the effects of poisoned samples during the training process to prevent backdoor creation; Model diagnosis \citep{tao2022better,guo2022aeva,xiang2022post} and sample diagnosis \citep{tran2018spectral,gao2021design,guo2023scale} try to detect whether a given suspicious model and sample is malicious, respectively; Different from previous types of defenses whose performances are empirical, certified defenses \citep{weber2022rab,jia2022certified,zeng2023towards} adopt randomized smoothing \citep{rosenfeld2020Certified} or linear bound propagation \citep{xu2020automatic} to certify the backdoor robustness of a given model under some conditions and assumptions. However, certified defenses usually suffer from low effectiveness and efficiency in practice \citep{li2021backdoor} since their assumptions do not hold in real-world situations.

Currently, our toolbox has implemented ten classical and advanced defenses, as shown in Table \ref{tab:defense}. We will keep updating this toolbox to include more representative defenses.

\begin{table}[!t]
\centering
\vspace{-2em}
\caption{Implemented backdoor defenses in \texttt{BackdoorBox}.}
\vspace{-1em}
\scalebox{0.95}{
\begin{tabular}{l|l|l}
\toprule
\textbf{Defense Type}                          & \textbf{Defense Method} & \textbf{Defender's Capacity}                                            \\ \hline
\multirow{2}{*}{Pre-processing-based} & AutoEncoder \citep{liu2017neural}    & \tabincell{l}{Black-box Model Accessibility\\  Samples for Training Auto-Encoder} \\ \cline{2-3}
                                      & ShrinkPad \citep{li2021backdoor}      & Black-box Model Accessibility                                    \\ \cline{2-3}\hline
\multirow{4}{*}{Model Repairing}      & Fine-tuning \citep{liu2018fine}    & \tabincell{l}{White-box Model Accessibility\\ Local Benign Samples}              \\ \cline{2-3}
                                      & Pruning \citep{liu2018fine}        & \tabincell{l}{White-box Model Accessibility\\ Local Benign Samples}               \\ \cline{2-3}
                                      & MCR \citep{zhao2020bridging}           & \tabincell{l}{White-box Model Accessibility\\ Local Benign Samples}               \\ \cline{2-3}
                                      & NAD \citep{li2021neural}           & \tabincell{l}{White-box Model Accessibility\\ Local Benign Samples}               \\ \hline
\multirow{3}{*}{Poison Suppression}   & ABL \citep{li2021anti}           & Training from Scratch                                            \\ \cline{2-3}
                                      & CutMix \citep{borgnia2021strong}        & Training from Scratch               \\ \cline{2-3}
                                      & DBD \citep{huang2022backdoor}           & Training from Scratch             \\ \hline
Sample Diagnosis                      & SS \citep{tran2018spectral}       & Obtaining Suspicious Dataset                                            \\ \bottomrule
\end{tabular}
}
\vspace{-1em}
\label{tab:defense}
\end{table}

\subsection{Library Design and Implementation}

\noindent \textbf{Design.} Similar to the design of attack module, all defense methods of the same type have the same core functions, consistent variable names, and the interface. Specifically, for pre-processing-based defenses, users can \textbf{(1)} adopt \texttt{preprocess} to obtain pre-processed data, \textbf{(2)} exploit \texttt{predict} to get predictions of the given (suspicious) model of pre-processed samples, and \textbf{(3)} \texttt{test} the performance of the defense on given datasets; For model repairing, users can \textbf{(1)} \texttt{repair} the given (suspicious) model, \textbf{(2)} obtained repaired model via \texttt{get\_model}, and \textbf{(3)} \texttt{test} the performance of the defense on given datasets; For poison suppression, users can \textbf{(1)} \texttt{train} a benign model based on suspicious samples, \textbf{(2)} obtain the trained model via \texttt{get\_model}, and \textbf{(3)} \texttt{test} its performance; For sample diagnosis, users can \texttt{filter} malicious samples and \texttt{test} the detection performance.

\vspace{0.3em}
\noindent \textbf{Implementation.} Please refer to Appendix \ref{appendix:defenses} for the demo examples for different types of defenses and the \href{https://github.com/THUYimingLi/BackdoorBox/blob/main/tests}{tests} directory for the test file of each method.


\section{Conclusion}
This paper presented \texttt{BackdoorBox}, a comprehensive and open-sourced Python toolbox for backdoor attacks and defenses. The consistency, simplicity, and flexibility of our toolbox make it an easy tool for researchers and developers to implement and develop, while the co-development ensures continuous updating. As avenues for future work, we plan to enhance our toolbox by implementing more advanced methods, improving its computational efficiency, supporting pip services, and developing methods towards other tasks and paradigms ($e.g.$, NLP and federated learning).

\section*{Acknowledgement}
This work is supported in part by the National Natural Science Foundation of China (62171248), the PCNL Key Project (PCL2021A07), the Shenzhen Philosophical and Social Science Plan (SZ2020D009), the PCNL Key Project (PCL2021A07), and the Tencent Rhino-Bird Research Program. We also sincerely thank Guanhao Gan, Kuofeng Gao, Jia Xu, Tong Xu, Sheng Yang, Xin Yan, Haoxiang Zhong, and Linghui Zhu, from Tsinghua Shenzhen International Graduate School (SIGS), Tsinghua University, for their assistance in co-developing and testing this toolbox.

\bibliography{iclr2023_conference}
\bibliographystyle{iclr2023_conference}

\newpage
\appendix
\section{The Demo Examples for Attacks}
In this section, we provide the code examples of using both poison-only attacks and training-controlled attacks in our \texttt{BackdoorBox}.

\subsection{The Example of Using BadNets as the Poison-only Attack}
\label{appendix:badnets_demo}

\renewcommand{\lstlistingname}{Demo}
\begin{lstlisting}[caption={The code example of using BadNets.}]
import torch
import torch.nn as nn
import core

# Assign the trigger pattern and its weight 
pattern = torch.zeros((32, 32), dtype=torch.uint8)
pattern[-3:, -3:] = 255
weight = torch.zeros((32, 32), dtype=torch.float32)
weight[-3:, -3:] = 1.0

# Initialize BadNets with adversary-specified hyper-parameters
badnets = core.BadNets(
    train_dataset=trainset, # Users should adopt their training dataset.
    test_dataset=testset, # Users should adopt their testing dataset.
    model=core.models.ResNet(18), # Users can adopt their model.
    loss=nn.CrossEntropyLoss(), 
    y_target=1, 
    poisoned_rate=0.05,
    pattern=pattern,
    weight=weight,
    deterministic=True
)

# Obtain the poisoned training and testing datasets
poisoned_train, poisoned_test = badnets.get_poisoned_dataset() 

# Train and obtain the attacked model
schedule = {
    'device': 'GPU',
    'CUDA_VISIBLE_DEVICES': '0',
    'GPU_num': 1,

    'benign_training': False,
    'batch_size': 128,
    'num_workers': 2,

    'lr': 0.1,
    'momentum': 0.9,
    'weight_decay': 5e-4,
    'gamma': 0.1,
    'schedule': [150, 180],

    'epochs': 200,

    'log_iteration_interval': 100,
    'test_epoch_interval': 10,
    'save_epoch_interval': 20,

    'save_dir': 'experiments',
    'experiment_name': 'ResNet-18_BadNets'
}

badnets.train(schedule) # Attack via given training schedule.
attacked_model = badnets.get_model() # Get the attacked model.
\end{lstlisting}

\subsection{The Example of Using PhysicalBA as the Training-controlled Attack}
\label{appendix:physicalba_demo}

\begin{lstlisting}[caption={The code example of using physical backdoor attack (PhysicalBA).}]
import torch
import torch.nn as nn
import core
from torchvision.transforms import Compose, ToTensor, PILToTensor, RandomHorizontalFlip, ColorJitter, RandomAffine

# Assign the trigger pattern and its weight 
pattern = torch.zeros((32, 32), dtype=torch.uint8)
pattern[-3:, -3:] = 255
weight = torch.zeros((32, 32), dtype=torch.float32)
weight[-3:, -3:] = 1.0

# Initialize PhysicalBA with adversary-specified hyper-parameters
PhysicalBA = core.PhysicalBA(
    train_dataset=trainset, # Users should adopt their training dataset.
    test_dataset=testset, # Users should adopt their testing dataset.
    model=core.models.ResNet(18), # Users can adopt their model.
    loss=nn.CrossEntropyLoss(), 
    y_target=1, 
    poisoned_rate=0.05,
    pattern=pattern,
    weight=weight,
    deterministic=True,
    physical_transformations = Compose([
        ColorJitter(brightness=0.2,contrast=0.2), 
        RandomAffine(degrees=10, translate=(0.1, 0.1), scale=(0.8, 0.9))
    ])    
)

# Train and obtain the attacked model
schedule = {
    'device': 'GPU',
    'CUDA_VISIBLE_DEVICES': '0',
    'GPU_num': 1,

    'benign_training': False,
    'batch_size': 128,
    'num_workers': 2,

    'lr': 0.1,
    'momentum': 0.9,
    'weight_decay': 5e-4,
    'gamma': 0.1,
    'schedule': [150, 180],

    'epochs': 200,

    'log_iteration_interval': 100,
    'test_epoch_interval': 10,
    'save_epoch_interval': 20,

    'save_dir': 'experiments',
    'experiment_name': 'ResNet-18_PhysicalBA'
}

PhysicalBA.train(schedule) # Attack via given training schedule.
attacked_model = PhysicalBA.get_model() # Get the attacked model.

# Obtain the poisoned training and testing datasets
poisoned_train, poisoned_test = PhysicalBA.get_poisoned_dataset() 
\end{lstlisting}

\section{The Demo Examples for Defenses}
\label{appendix:defenses}

In this section, we provide the code examples of using pre-processing-based defenses, model repairing, poison suppression, and sample diagnosis in our \texttt{BackdoorBox}.

\subsection{The Example of Using ShrinkPad as the Pre-processing-based Defense}

\begin{lstlisting}[caption={The code example of using ShrinkPad.}]
import torch
import torch.nn as nn
import core

# Initialize ShrinkPad with defender-specified hyper-parameters
ShrinkPad = core.ShrinkPad(
    size_map=32, # Users should assign it based on their samples.
    pad=4, # Key hyper-parameter of ShrinkPad.
    deterministic=True
)

# Get the pre-processed images
pre_img = ShrinkPad.preprocess(img) # Users should use their images.

# Get the predictions of pre-processed images by the given model
predicts = ShrinkPad.predict(model, img)

# Define the test schedule
schedule = {
    'device': 'GPU',
    'CUDA_VISIBLE_DEVICES': '0',
    'GPU_num': 1,
    
    'batch_size': 128,
    'num_workers': 2,

    'metric': 'ASR_NoTarget',
    'y_target': y_target,

    'save_dir': 'experiments',
    'experiment_name': 'ShrinkPad-4_ASR_NoTarget'
}

# Evaluate the performance of ShrinkPad on a given dataset
ShrinkPad.test(model, dataset, schedule)


\end{lstlisting}

\subsection{The Example of Using Fine-tuning as the Model Repairing}

\begin{lstlisting}[caption={The code example of using fine-tuning.}]
import torch
import torch.nn as nn
import core

# Initialize fine-tuning with defender-specified hyper-parameters
finetuning = core.FineTuning(
    train_dataset=dataset, # Users should adopt their benign samples.
    test_dataset=dataset_test # Users can use both benign and poisoned datasets for evaluation.
    model=model, # Users should adopt their suspicious model.
    layer=["full layers"], # Users should assign their tuning position.
    loss=nn.CrossEntropyLoss(),
)

# Define the repairing schedule 
schedule = {
    'device': 'GPU',
    'CUDA_VISIBLE_DEVICES': '0',
    'GPU_num': 1,

    'batch_size': 128,
    'num_workers': 4,

    'lr': 0.001,
    'momentum': 0.9,
    'weight_decay': 5e-4,
    'gamma': 0.1,

    'epochs': 10,
    'log_iteration_interval': 100,
    'save_epoch_interval': 2,

    'save_dir': 'experiments',
    'experiment_name': 'finetuning'
}

# Repair the suspicious model
finetuning.repair(schedule)

# Obtain the repaired model
repaired_model = finetuning.get_model()

# Evaluate the performance of repaired model with given testing schedule
test_schedule = {
    'device': 'GPU',
    'CUDA_VISIBLE_DEVICES': '0',
    'GPU_num': 1,

    'batch_size': 128,
    'num_workers': 4,
    'metric': 'BA',

    'save_dir': 'experiments',
    'experiment_name': 'finetuning_BA'
}

finetuning.test(benign_dataset, test_schedule)
\end{lstlisting}

\subsection{The Example of Using CutMix as the Poison Suppression}

\begin{lstlisting}[caption={The code example of using CutMix.}]
import torch
import torch.nn as nn
import core

# Initialize CutMix with defender-specified hyper-parameters
CutMix = core.CutMix(
    model=model, # Users should adopt their model
    loss=nn.CrossEntropyLoss(),
    beta=1.0,
    cutmix_prob=1.0,
    deterministic=True
)


# Train the model with a given schedule
schedule = {
    'device': 'GPU',
    'CUDA_VISIBLE_DEVICES': '0',
    'GPU_num': 1,

    'batch_size': 128,
    'num_workers': 4,

    'lr': 0.1,
    'momentum': 0.9,
    'weight_decay': 5e-4,
    'gamma': 0.1,
    'schedule': [150, 180],

    'epochs': 200,

    'log_iteration_interval': 100,
    'test_epoch_interval': 20,
    'save_epoch_interval': 20,

    'save_dir': 'experiments',
    'experiment_name': 'CutMix',
}

CutMix.train(trainset=trainset, schedule=schedule) # Users should adopt their local suspicious training dataset.

# Obtain the trained model 
model = CutMix.get_model()

# Evaluate the performance of trained model with given testing schedule
test_schedule = {
    'device': 'GPU',
    'CUDA_VISIBLE_DEVICES': '0',
    'GPU_num': 1,

    'batch_size': 128,
    'num_workers': 4,
    'metric': 'BA',

    'save_dir': 'experiments',
    'experiment_name': 'CutMix_BA'
}

CutMix.test(benign_dataset, test_schedule)
\end{lstlisting}

\subsection{The Example of Using SS as the Sample Diagnosis}

\begin{lstlisting}[caption={The code example of using spectral signature (SS).}]
import torch
import torch.nn as nn
import core

# Initialize SS with defender-specified hyper-parameters
Spectral = core.SS(
    model=model, # Users should adopt the model trained on suspicious dataset.
    dataset=suspicious_dataset, 
    percentile=80, # Key hyper-parameter of SS.
    deterministic=True
)

# Filter out poisoned samples
poisoned_idx, _ = Spectral.filter()

# Evaluate the performance of SS with given testing schedule
test_schedule = {
    'device': 'GPU',
    'CUDA_VISIBLE_DEVICES': '0',
    'GPU_num': 1,

    'batch_size': 128,
    'num_workers': 4,
    'metric': 'Precision',

    'save_dir': 'experiments',
    'experiment_name': 'SS_Precision'
}

Spectral.test(poisoned_idx_true, test_schedule)


\end{lstlisting}

\section{The Toolbox Structure}
\begin{figure*}[ht]
    \centering
    \includegraphics[width=\textwidth]{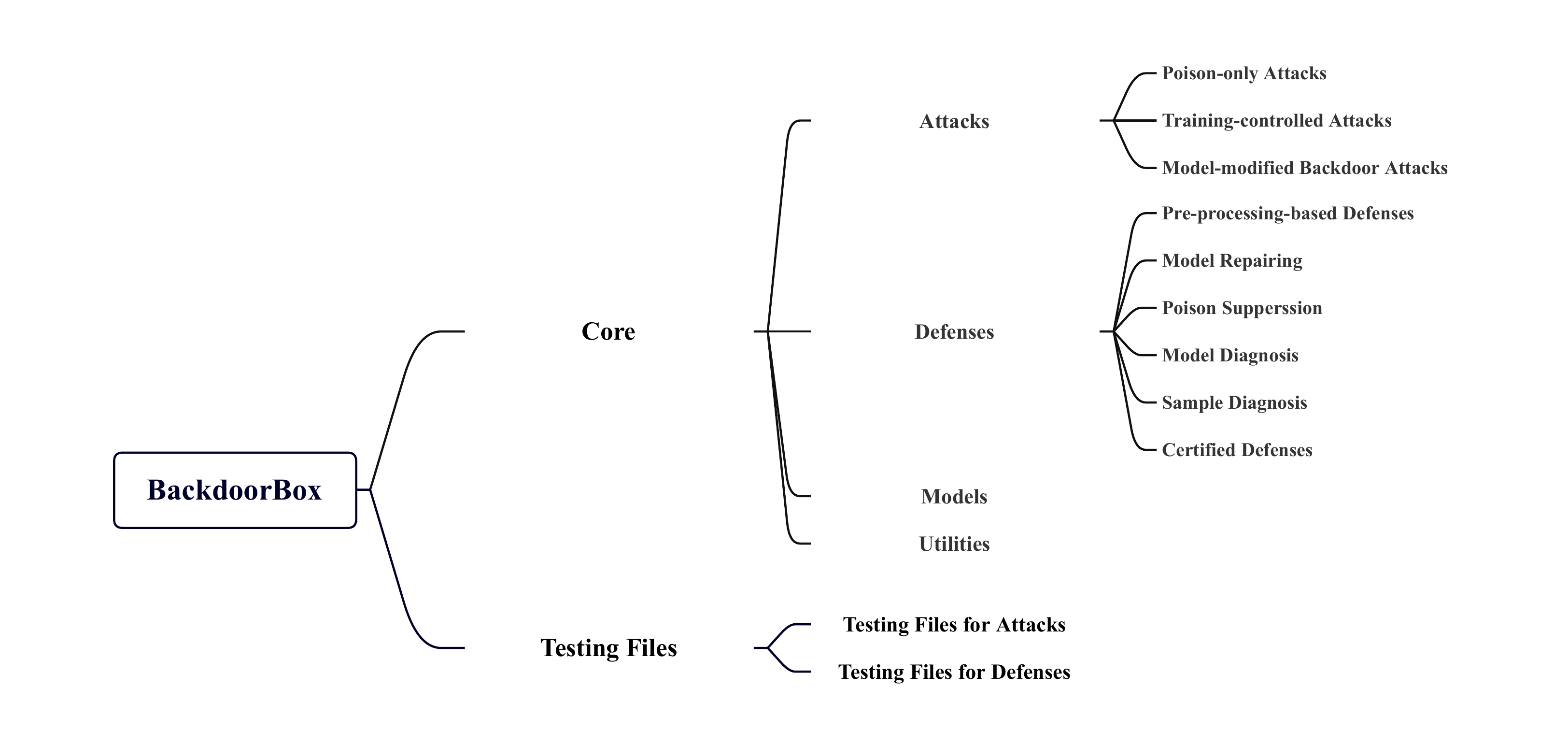}
    \vspace{-1em}
    \caption{The framework of our \texttt{BackdoorBox}.}
    \label{fig:framework}
\end{figure*}

As shown in Figure \ref{fig:framework}, our toolbox consists of five main parts, including \textbf{(1)} \href{https://github.com/THUYimingLi/BackdoorBox/tree/main/core/attacks}{attack module}, \textbf{(2)} \href{https://github.com/THUYimingLi/BackdoorBox/tree/main/core/defenses}{defense module}, \textbf{(3)} \href{https://github.com/THUYimingLi/BackdoorBox/tree/main/core/models}{model module}, \textbf{(4)} \href{https://github.com/THUYimingLi/BackdoorBox/tree/main/core/utils}{utility module}, and \textbf{(5)} \href{https://github.com/THUYimingLi/BackdoorBox/tree/main/tests}{testing files}. The first four parts are the core functional components of our toolbox, while the last one is provided for users as code examples. Specifically, the attack and defense modules contain all implemented attacks and defenses where each method is in a separate Python file; The model module contains classical model architectures. Users can freely adopt these provided models or their local models; The utility module provides supportive functionalities, such as the calculation of evaluation metrics; The code example of each implemented method is included in testing files.

Currently, our toolbox has only implemented part of both attack and defense modules (as suggested in Section \ref{sec:attacks}-\ref{sec:defenses}). We will keep updating this toolbox and developing representative methods and necessary supportive components that have not been included.

\end{document}